\documentclass[
reprint,
superscriptaddress,
amsmath,
amssymb,
aps,
prb,
floatfix,
]{revtex4-2}
\usepackage{graphicx}
\usepackage{hyperref}
\usepackage{xcolor}
\usepackage{multirow}

\begin{document}

\title{Lattice vacancy migration barriers in  Fe-Ni alloys, \\ and why Ni atoms diffuse slowly: An \textit{ab initio} study}

\author{Adam M. Fisher}
\thanks{Present address: United Kingdom Atomic Energy Authority, Culham Campus, Oxfordshire, OX14 3DB, United Kingdom}
\affiliation{Department of Physics, University of Warwick, Coventry, CV4 7AL, United Kingdom}
\author{Christopher D. Woodgate}
\thanks{Corresponding author}
\email{christopher.woodgate@bristol.ac.uk}
\affiliation{H.H. Wills Physics Laboratory, University of Bristol, Royal Fort, Bristol, BS8 1TL, United Kingdom}
\author{Xiaoyu Zhang}
\affiliation{Department of Chemical Engineering, Northeastern University, Boston, MA 02115, USA}
\affiliation{Department of Mechanical and Industrial Engineering, Northeastern University, Boston, MA 02115, USA}
\author{George C. Hadjipanayis}
\affiliation{Department of Chemical Engineering, Northeastern University, Boston, MA 02115, USA}
\author{Laura H. Lewis}
\affiliation{Department of Chemical Engineering, Northeastern University, Boston, MA 02115, USA}
\affiliation{Department of Mechanical and Industrial Engineering, Northeastern University, Boston, MA 02115, USA}
\affiliation{Department of Physics, Northeastern University, Boston, MA 02115, USA}
\author{Julie B. Staunton}
\email{J.B.Staunton@warwick.ac.uk}
\affiliation{Department of Physics, University of Warwick, Coventry, CV4 7AL, United Kingdom}

\begin{abstract}
{ Lattice vacancy migration barriers} in ferromagnetic Fe$_x$Ni$_{1-x}$ alloys ($0.4 \leq x \leq 0.6$) are { accurately quantified} within the framework of \textit{ab initio} electronic structure calculations using the nudged elastic band (NEB) method.
Both the atomically disordered (A1) fcc phase, as well as the atomically ordered, tetragonal L1\textsubscript{0} phase---which is under consideration as a material for a rare-earth-free `gap' magnet for advanced engineering applications---are investigated.
Across an ensemble of NEB calculations performed on supercell configurations spanning a range of compositions and containing disordered, partially ordered, and fully ordered structures, we find that Ni{-vacancy interchanges encounter significantly higher energetic barriers than do Fe-vacancy interchanges.}
{We contend that this aspect is a key factor in determining the differences in mobility between Fe and Ni atoms in this ferromagnetic alloy.}
{Moreover}, we are able to interpret these findings in terms of the ferromagnetic alloy's underlying spin-polarised electronic structure.
Specifically, we report a coupling between the size of local lattice distortions and the magnitude of the local electronic spin polarisation around vacancies.
This causes Fe atoms to relax into lattice vacancies, while Ni atoms remain rigidly fixed to their original lattice positions. 
These results give atomic-scale insight into the longstanding experimental observation that Ni exhibits remarkably slow atomic diffusion in Fe-Ni alloys.
\end{abstract}

\date{May 10, 2026}

\maketitle

\section{Introduction}
\label{sec:introduction}

A lattice vacancy is a point defect characterized by the absence of an atom occupying a particular crystal lattice site. At any finite temperature,  lattice vacancies are present at a given equilibrium concentration, with the relative concentration of lattice vacancies increasing with increasing temperature~\cite{callister_materials_2020}. The diffusivity of  constituent elements in alloys impacts metallurgical reactions such as chemical ordering, segregation, and precipitation and proceeds primarily via vacancy-mediated atomic diffusion, where lattice vacancies interchange with atoms occupying neighbouring lattice sites~\cite{huntington_mechanism_1942, huntington_self-consistent_1942}. 

The rates of diffusion of solute elements in alloys---particularly those containing Fe---are known to differ substantially~\cite{bowen_solute_1970, oikawa_lattice_1982}. For example, in austenitic Fe-Cr-Ni alloys, across a range of temperatures and compositions, it is consistently found that Cr has a higher diffusion coefficient than that of Fe, which in turn has a higher diffusion coefficient than that of Ni~\cite{rothman_self-diffusion_1980}. Another example is found in the complex concentrated Cr-Mn-Fe-Co-Ni system, where Mn atoms are most mobile, followed (in order) by Cr, Fe, Co, through to Ni, which is least mobile~\cite{tsai_sluggish_2013}. However, atomic-scale insight into the underlying physical origins of these varying diffusivities is lacking in many cases and warrants further investigation.

\begin{figure}[t]
    \centering
    \includegraphics[width=\linewidth]{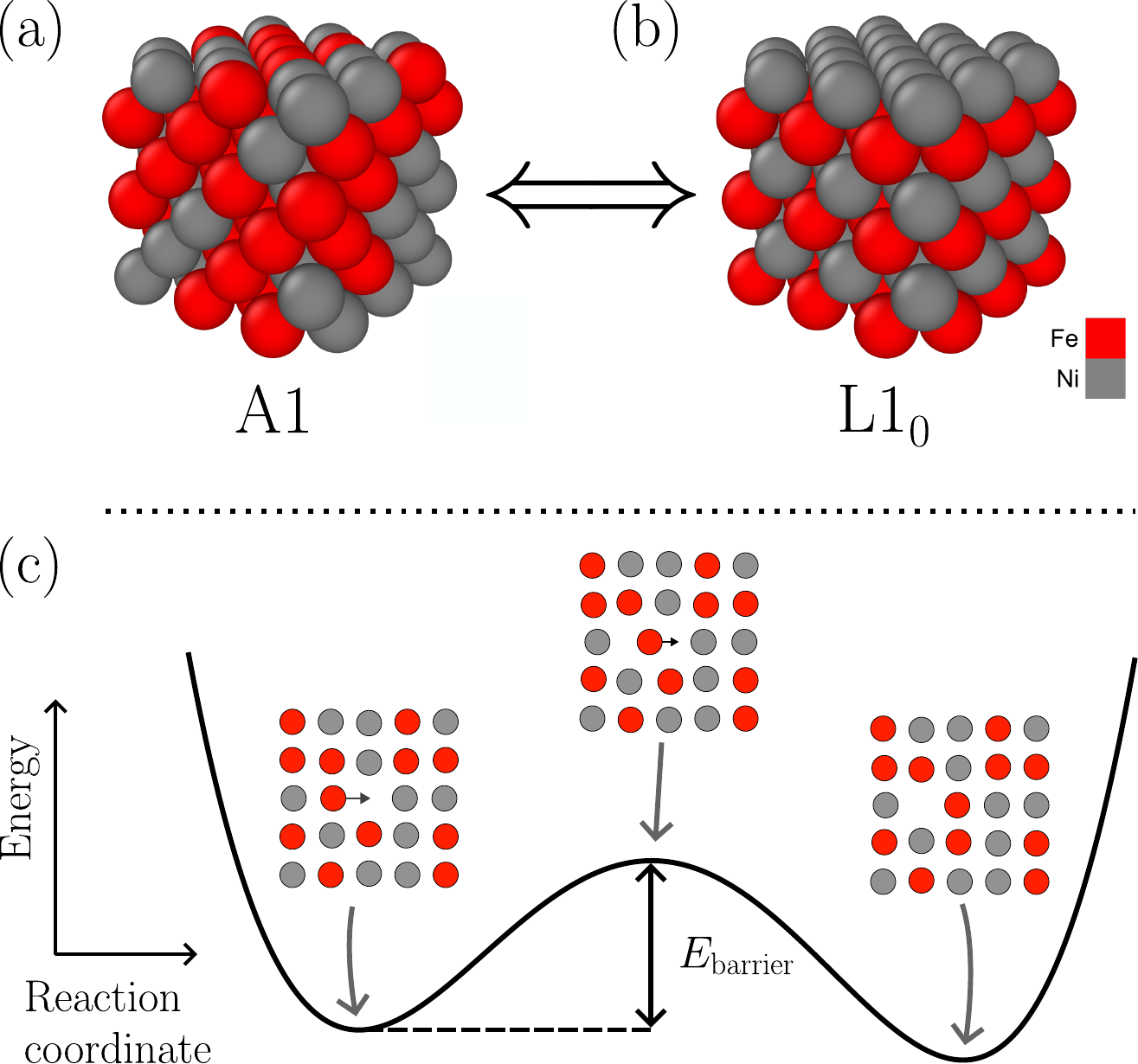}
    \caption{A conceptual illustration of the focus of the present study. The transformation from the A1 structure (a) to the L1\textsubscript{0} structure (b) in Fe-Ni alloys is mediated by lattice vacancy migration (c). In this context, the A1 structure represents an atomically disordered fcc system, while the L1\textsubscript{0} structure represents a tetragonal, atomically ordered system, with desirable associated hard magnetic properties. The `height' of a lattice vacancy migration barrier, denoted $E_\textrm{barrier}$, is related to how much thermal energy input is required to interchange an atom with a neighbouring vacancy. The images in panels (a) and (b) were generated using \textsc{Ovito}~\cite{stukowski_visualization_2010}.}\label{fig:ordering_barrier_schematic}
\end{figure}

One alloy system where atomic diffusion rates are of particular interest is the Fe-Ni binary system, where several intermetallic phases are known to emerge at comparatively low temperatures~\cite{von_goldbeck_ironnickel_1982, howald_thermodynamics_2003}. In particular, around the equiatomic composition, it is known that it is thermodynamically favourable for the material to crystallize into a tetragonal, atomically ordered structure; Strukturbericht designation L1\textsubscript{0}, space group $P4/mmm$. This ordered L1\textsubscript{0} phase consists of layers of Fe and Ni atoms alternating on an underlying face-centred tetragonal lattice~\cite{mehl_aflow_2017}, and is illustrated in Fig.~\ref{fig:ordering_barrier_schematic}. It was first synthesised by N\'eel, Paulev\'e, and co-workers~\cite{pauleve_nouvelle_1962, neel_magnetic_1964, pauleve_magnetization_1968}, and was also later discovered to occur naturally in very slowly-cooled metallic meteorites, with typical cooling rates for such meteorites determined to be in the range 0.3~--~6600~K~Myr$^{-1}$~\cite{yang_low-temperature_1997, lewis_inspired_2014}. This ordered phase is therefore sometimes referred to by its meteoritic mineral name: tetrataenite~\cite{clarke_tetrataeniteordered_1980}. L1\textsubscript{0} FeNi is known, both from experimental measurements~\cite{shima_structure_2007, mizuguchi_artificial_2011, kojima_l10-ordered_2011, kojima_magnetic_2012, kojima_feni_2014, poirier_intrinsic_2015, frisk_resonant_2016, frisk_strain_2017, ito_fabrication_2023} and theoretical calculations~\cite{miura_origin_2013, edstrom_electronic_2014, lewis_magnete_2014, werwinski_ab_2017, izardar_interplay_2020, si_effect_2022, yamashita_first-principles_2022, woodgate_revisiting_2023, yamashita_finite-temperature_2023, marciniak_magnetic_2024}, to possess a sufficiently large uniaxial magnetocrystalline anisotropy energy to be useful as a rare-earth-free `gap' magnet~\cite{coey_permanent_2012}, with an anticipated maximum energy product between that of the rare-earth supermagnets and the oxide ferrites~\cite{lewis_inspired_2014}.

However, although thermodynamically stable relative to its precursor atomically disordered face-centred cubic (fcc) phase (the meteoritic mineral known as taenite), the atomically ordered L1\textsubscript{0} phase remains challenging to synthesise in the laboratory. This difficulty  arises from the low atomic ordering temperature and highly sluggish diffusion kinetics~\cite{dos_santos_kinetics_2015, bordeaux_thermodynamic_2016}. Attempts to achieve bulk synthesis of L1\textsubscript{0} FeNi have therefore focused on enhancing atomic diffusion, for example by introducing lattice defects that facilitate atomic mobility~\cite{mandal_l10_2023, chamberod_electron_1979, lee_formation_2014, geng_defect_2015, maat_creating_2020, lewis_accelerating_2023, mandal_formation_2023}. Additionally, it is understood that the thermodynamic stability of the L1\textsubscript{0} phase is improved when the system is in its ferromagnetic state~\cite{izardar_interplay_2020, ruban_qualitative_2024, woodgate_integrated_2024}. As a result, processing must take place below the Curie temperature of the disordered phase, imposing an additional constraint that can limit the rate of phase transformation. 

Atomic diffusion in Fe-Ni alloys has been extensively experimentally investigated~\cite{macewan_diffusion_1959, hirano_diffusion_1961, de_reca_self-diffusion_1967, bakker_curvature_1971, ustad_interdiffusion_1973, million_diffusion_1981, narayan_low_1983, ganesan_interdiffusion_1984,  dean_determination_1986, yang_magnetic_2004}, and it is concluded that Ni atoms are significantly less mobile than Fe atoms~\cite{macewan_diffusion_1959, de_reca_self-diffusion_1967, million_diffusion_1981}. This sluggish diffusion of Ni atoms { likely} reflects, { in part}, the relatively high energy barrier for atomic motion---defined as the energy input required for an atom to move to a neighbouring lattice site (typically a vacancy, illustrated in Fig.~\ref{fig:ordering_barrier_schematic}). Therefore, understanding and overcoming the high energy barrier to Ni diffusion emerges as a key challenge for realizing bulk synthesis of L1\textsubscript{0} FeNi for advanced technological applications.

Atomic-scale diffusion processes in Fe-Ni alloys have also been the subject of several previous computational studies~\cite{zhao_defect_2016, anento_effect_2017, mahmoud_long-time_2018, osetsky_tunable_2020, li_predicting_2023} { considering the mobility of} both { lattice} vacancies~\cite{zhao_defect_2016, anento_effect_2017, mahmoud_long-time_2018, li_predicting_2023} and self-interstitials~\cite{zhao_defect_2016, anento_effect_2017, mahmoud_long-time_2018, osetsky_tunable_2020}. In these studies, the potential energy landscape is described using both semi-empirical potentials~\cite{zhao_defect_2016, anento_effect_2017, mahmoud_long-time_2018, li_predicting_2023} and \textit{ab initio} calculations~\cite{zhao_defect_2016}. It is generally found that Ni atoms are less mobile than Fe, consistent with experimental observations. Additionally, previous semi-empirical work has suggested that magnetic order plays a key role in determining interdiffusion coefficients in both bcc and fcc Fe-Ni alloys, with magnetic order understood to raise the energy barrier for atomic motion, leading to suppressed diffusion in Fe-Ni systems~\cite{yang_magnetic_2004}. However, a clear, atomic-scale explanation for these observed differences in atomic mobility between Ni and Fe atoms in ferromagnetic Fe-Ni alloys has yet to be developed. 

To that end, here we { shed light on aspects of} the underlying atomic-scale physical mechanism driving the difference in  mobility between Fe and Ni atoms by considering the changes to the local structure and magnetism of the environment surrounding the lattice vacancies as compared to the bulk material. This is approached through application of \textit{ab initio} density functional theory (DFT) calculations, as previous works have shown that semi-empirical potentials often fail to accurately capture lattice vacancy migration barrier heights in metallic alloys~\cite{zhao_defect_2016, fisher_first_2024}. We find that, across an ensemble of lattice vacancy migration barriers evaluated using the nudged elastic band (NEB) method~\cite{Jonsson1998}, the height of energy barriers associated with Ni-vacancy interchanges is consistently significantly higher than those associated with Fe-vacancy interchanges. { We contend that this} outcome is in alignment with experimental findings { concerning species-dependent atomic mobility in these alloys}. Proceeding, examination of local changes to the alloy's spin-polarised electronic structure in the vicinity of lattice vacancies reveals that Ni atoms remain rigidly fixed to their original lattice positions, while Fe atoms `relax' significantly into lattice vacancies. This displacement of Fe atoms is associated with an increased magnitude of spin polarisation (local magnetic moment) on Fe atoms adjacent to these vacancies. Subsequent analysis confirms that this effect is the origin of the difference in average lattice vacancy migration barrier height (related to the activation energy for diffusion) for Fe-vacancy interchanges as opposed to Ni-vacancy interchanges. { We believe that} these results { go some way towards providing} an intuitive physical explanation for long-standing experimental observations that { Ni atoms are significantly less mobile than Fe atoms in ferromagnetic, fcc} Fe-Ni alloys.

The remainder of this paper is structured as follows. In Sec.~\ref{sec:methods}, the details of our computational modelling and methodology for studying lattice vacancy migration barriers are outlined, including discussion of how a range of potential chemical compositions and degrees of atomic ordering are taken into account. In Sec.~\ref{sec:results}, results are presented, where it is found that the energetic barriers associated with Ni atom-vacancy interchanges are significantly higher than those associated with Fe atom-vacancy interchanges.  Fundamental insight into this outcome is extracted by examination of both local structural changes around the lattice vacancies in these alloys as well as investigation of induced changes to the electronic structure. Finally, in Sec.~\ref{sec:conclusions}, the findings are summarised, and potential future avenues of study are suggested.

\section{Methods}
\label{sec:methods}

\begin{figure*}
    \centering
    \includegraphics[width=\linewidth]{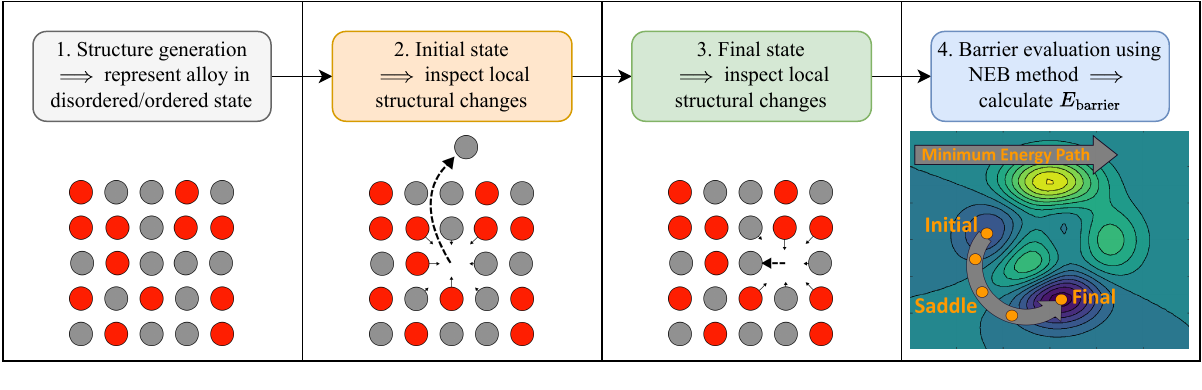}
    \caption{An illustration of the workflow used in this study for calculation of lattice vacancy migration barriers in disordered and ordered Fe-Ni alloys using the NEB method. These stages are discussed in more detail in Sec.~\ref{sec:methods}.}
    \label{fig:workflow}
\end{figure*}

In this work, { lattice vacancy migration barriers} in both disordered and ordered, near-equiatomic Fe-Ni alloys, { are} investigated using \textit{ab initio} electronic structure calculations performed on alloy supercells. The NEB method~\cite{Jonsson1998}, which is a well-established method for searching for transition paths, facilitates accurate evaluation of lattice vacancy migration barriers. Moreover, analysis of the extensive, \textit{ab initio} dataset reveals changes to the local atomic structure around lattice vacancies, which is interpreted in terms of the alloys' underlying electronic and magnetic structures and, in turn, results in { differing lattice vacancy migration barriers for Ni-vacancy interchanges as compared to Fe-vacancy interchanges}.

The workflow for performing a NEB calculation on an alloy supercell for calculation of a lattice vacancy migration barrier, illustrated in Fig.~\ref{fig:workflow}, has four stages and is summarised as follows:
\begin{enumerate}
    \item Generate an fcc supercell representative of the Fe-Ni alloy in either a fully atomically disordered, short-range ordered, or L1\textsubscript{0} ordered state.
    \item Remove a single atom---selected at random---from the simulation cell to create a lattice vacancy. Then perform a `geometry optimization' (or `relaxation') to determine the arrangement of atoms around the vacancy which minimizes the internal energy as calculated using DFT. This is the initial state (sometimes referred to as an `image') for the NEB calculations which are to follow. From these calculations, it is possible to recover both how the local, atomic-scale structure of the alloy evolves around lattice vacancies, as well as how changes to the alloy's magnetism and electron density drive these structural changes. 
    \item Select---at random---one of the twelve atoms neighbouring the lattice vacancy and interchange this atom into the vacancy, thus moving the lattice vacancy to a neighbouring lattice site. A geometry optimization is then performed on this interchanged structure to create the final state (or `image') for use in the NEB calculation.
    \item Perform a NEB calculation to determine the lattice vacancy migration barrier. Starting from an initial series of `images' along the path, the NEB method finds the minimum energy path connecting the initial and final state.
\end{enumerate}
We discuss details of generation of representative alloy structures, as well as computational details of our DFT and NEB calculations, below.

\subsection{Supercell generation}
\label{subsec:supercell_generation}

To study lattice vacancy migration barriers in the Fe$_x$Ni$_{1-x}$ ($0.4\leq x \leq 0.6$) alloys considered in this work, appropriate vacancy-containing supercells in both atomically disordered (A1), partially-ordered, and fully atomically ordered (L1\textsubscript{0}) phases are required across a range of values of $x$. { To generate the disordered and partially ordered structures, we follow the approach proposed by Shenoy \textit{et al.}~\cite{shenoy_collinear-spin_2024}, and draw samples from equilibrated, lattice-based Metropolis Monte Carlo~\cite{metropolis_equation_1953} simulations at a range of simulation temperatures. In these fixed-lattice simulations, performed using the \textsc{Brawl} package~\cite{naguszewski_brawl_2025}, the alloy internal energy is described as a sum over atom-atom effective pair interactions, which are obtained within an effective medium theory~\cite{khan_statistical_2016, woodgate_compositional_2022, woodgate_modelling_2024}.} Those simulation cells equilibrated at low temperatures will have a high degree of atomic order, while those equilibrated at higher temperatures will be more representative of the disordered phase. { Full details of this approach are discussed in Appendix~\ref{appendix:supercell}. For lattice vacancy migration barriers, 32-atom simulation cells (supercells) were used, with periodic boundary conditions applied in all three directions. (This comparatively small simulation cell, which is consistent with the size of simulation cell used in a previous \textit{ab initio} study on fcc Ni~\cite{hargather_first-principles_2014}, is necessary given the high computational cost of the NEB calculations for evaluation of lattice vacancy migration barriers.) For calculations of vacancy formation energies, larger, 108-atom simulation cells were employed. In addition to the disordered supercells,} we also construct 32- and 108-atom supercells representing the pristine, fully ordered L1\textsubscript{0} phase for the equiatomic composition and assess how lattice vacancies form and { migrate} in this phase. (There are four unique vacancy migration barriers that are possible in a perfectly ordered L1\textsubscript{0} FeNi system with a single isolated vacancy, \textit{i.e.} a vacancy in either the Fe or Ni layer and an atomic interchange occurring either in-layer or between layers.) Calculations performed on this pristine phase contribute to the understanding of the behaviour of vacancies in the ordered alloy.

\subsection{Calculation of lattice vacancy migration barrier heights}

A starting point for an initial barrier evaluation is set by randomly removing one atom from a particular supercell to create a vacancy. Proceeding, one of the 12 nearest-neighbouring atoms to the vacant site is randomly selected as the atom that moves into the vacancy, determining the final state of the energy barrier calculation. The supercell configurations associated with both initial and final states are relaxed to find the local minimum of the energy landscape. A fixed lattice parameter, corresponding to the average lattice parameter of the disordered solid solution, is employed throughout. A fixed lattice parameter is required to avoid significant distortion of the unit cell under large vacancy concentration, as large distortions of the unit cell are incompatible with subsequent vacancy migration energy barrier calculations which require maintenance of fixed simulation cell dimensions for all `images'.

Lattice vacancy migration barriers were then evaluated using the NEB method \cite{Jonsson1998} as implemented in CASTEP. We provide a conceptual illustration of this method in the rightmost pane of Fig.~\ref{fig:workflow}. We used the \texttt{ode12r} optimiser \cite{Makri2019}, and the climbing image NEB method \cite{Henkelman20009901} for reliable convergence to the saddle point. We tested convergence with respect to number of NEB images used by comparing barrier heights evaluated with 5, 7, and 9 images, and found that 5 images was adequate. (During testing, we found 5 images typically obtained a barrier height which was within 1~meV of the value obtained with 7 and 9 images, which we attribute to the reliability of the climbing image NEB method when converging to the saddle point.) In total, 192 lattice vacancy migration barriers were considered, 103 of which were Fe-vacancy interchanges, 89 of which were Ni-vacancy interchanges. { For each NEB calculation, there are two calculated barriers: the barrier associated with a vacancy migrating from the selected initial state to the selected final state (the `left' barrier), and the barrier associated with a vacancy migrating from the selected final state to the selected initial state (the `right' barrier). In the main text, all plots present combined data for left and right barriers, \textit{i.e.} 384 data points, while in the Supplemental Material~\cite{supplemental} we provide separate plots decomposed into data points for left and right barriers.}

We find that---in this close-packed (fcc) binary alloy where both alloying elements have similar atomic sizes---the evaluated barriers typically have a single maximum and a comparatively smooth energy profile as a function of reaction coordinate. Plots evidencing this smooth behaviour across the dataset are provided in the Supplemental Material~\cite{supplemental}. A select few barriers where the NEB calculation failed to converge were excluded from the analysis.

{ \subsection{Calculation of vacancy formation energies}}

{ In addition to the main focus of this study---calculation of lattice vacancy migration barrier heights---we also provide calculations of species-dependent vacancy formation energies in the alloy. Vacancies are understood to primarily originate from the surface(s) of a crystal~\cite{mccarty_vacancies_2001}. Though in the bulk we expect these vacancies to be long-lived and, thus, vacancy migration barriers to be the dominant factor determining atomic mobility, it is nonetheless important to consider the energetic cost of vacancy formation for the different chemical species present in the alloy for comparison with the evaluated lattice vacancy migration barriers.

In this work, the vacancy formation energy for a vacancy in chemical species $\alpha$, $E^\textrm{f}_\textrm{v}(\alpha)$, is calculated according to the formula~\cite{freysoldt_first-principles_2014}
\begin{equation}
E^\textrm{f}_\textrm{v}(\alpha) = E^\textrm{tot}_\textrm{vac}(\alpha) + \mu_\alpha - E^\textrm{tot}_\textrm{prist},
\label{eq:vacancy_formation}
\end{equation}
where $E^\textrm{tot}_\textrm{vac}(\alpha)$ is the total energy of a (relaxed) simulation cell where an atom of chemical species $\alpha$ has been removed to create the vacancy, $\mu_\alpha$ is the chemical potential associated with species $\alpha$, and $E^\textrm{tot}_\textrm{prist}$ is the total energy of the (relaxed) pristine supercell without the vacancy. We approximate $\mu_\alpha$ by taking this to be the energy per atom of the pure bulk element of each chemical species. For completeness, results are provided for $\mu_\textrm{Fe}$ taken to be the energy of bcc Fe and (metastable) fcc Fe. For evaluation of $E^\textrm{tot}_\textrm{vac}(\alpha)$, we perform an unconstrained geometry optimisation at zero pressure, meaning that we strictly calculate the enthalpy of vacancy formation.\\}

{
\subsection{Computational details}}

Geometry optimisations and NEB calculations were performed using the CASTEP package~\cite{Clark2005, castep_webpage}, which implements plane-wave pseudopotential DFT~\cite{payne_iterative_1992}. We used the generalized gradient approximation (GGA) to the exchange-correlation functional as parametrised by Perdew, Burke, and Ernzerhof (PBE)~\cite{perdew_generalized_1996}. To accurately converge barrier heights, it is necessary to use very precise DFT settings. Consequently, we used a plane-wave cutoff energy of 800 eV, with an electronic smearing width of 0.05 eV, corresponding to an electronic temperature of 580 K, \textit{i.e.} close to the ordering temperature of the alloy~\cite{pauleve_nouvelle_1962} as is relevant to this study. Core electronic states were described using ultrasoft pseudopotentials~\cite{vanderbilt_soft_1990, laasonen_car-parrinello_1993} generated `on-the-fly' within CASTEP. Brillouin zone integrations were performed using a Monkhorst-Pack grid~\cite{monkhorst_special_1976} with a maximum allowed $\mathbf{k}$-point spacing of 0.02~\AA$^{-1}$. (For the 32-atom supercells considered in this work, this results in a dense $8 \times 8 \times 8$ $\mathbf{k}$-point mesh over the first Brillouin zone.) An electronic self-consistency tolerance of $1\times10^{-7}$ eV was applied to the electronic portion of the calculations, while, for geometry optimizations, forces converged to a tolerance of $1 \times 10^{-4}$~eV/\AA\ and stresses to a tolerance of $1 \times 10^{-3}$~GPa. Using these settings, we recover lattice parameters for the pristine, equiatomic L1\textsubscript{0} phase of $a=b=3.5543$~\AA, $c = 3.5796$~\AA, and $c/a =  1.0071$, in good agreement with the experimentally determined values~\cite{albertsen_tetragonal_1981} of $a = b = 3.5761 \pm 0.0005$~\AA, $c = 3.5890 \pm 0.0005$~\AA, and $c/a = 1.0036 \pm 0.0002$~\AA. For the disordered, A1 phase (represented using the simulation cells equilibrated at a high simulation temperature of 3000 K), the average fcc lattice parameter is $3.5687$~\AA. Further details of our calculations, including all relevant CASTEP input and output files, can be found in the open-access dataset associated with this study~\cite{dataset}.

Generation of initial CASTEP input files from the equilibrated Monte Carlo simulations, and analysis of relevant outputs, were performed using the interface to CASTEP provided by the Atomic Simulation Environment (ASE)~\cite{hjorth_larsen_atomic_2017}. Where calculations of such quantities as interatomic distances and/or bond lengths are considered, periodic boundary conditions and the minimum image convention were applied throughout. { Where atomic magnetic moments are presented, these are calculated using the Mulliken population analysis~\cite{mulliken_electronic_1955, segall_population_1996, segall_population_1996-1} as implemented in \textsc{CASTEP}, which projects the final self-consistent electron density onto a local basis of atomic orbitals. In contrast, where the total magnetisation of a cell is presented, this is obtained by evaluating the difference between spin up and spin down density across the simulation cell. In all calculations presented in this study, the charge spilling parameter for the Mulliken analysis (quantifying how `good' the projection onto the local basis is for population analysis) is small, typically 0.1--0.2\%, and we are therefore confident that the atomically-resolved magnetic moments are reliable.\\}

\section{Results and Discussion}
\label{sec:results}

\subsection{Lattice vacancy migration barriers}
\label{sec:results_barriers}

\begin{figure}[b]
    \centering
	\includegraphics[width=\linewidth]{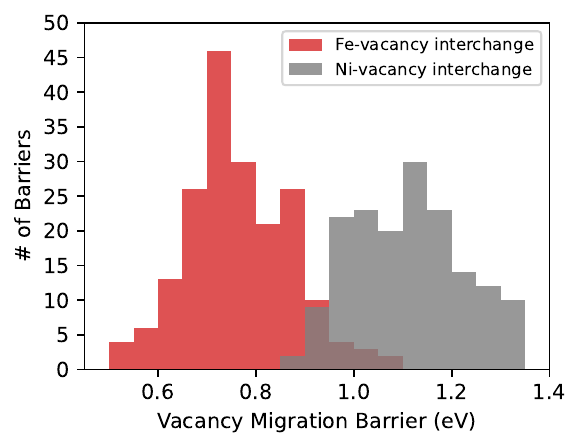}
	\caption{Histograms of lattice vacancy migration barriers across the ensemble of considered Fe-Ni alloy supercell configurations in this study. Results are compared for interchanges where Fe atoms are switched with vacancies (red) versus interchanges where Ni atoms are switched into  vacancies (grey). Across the ensemble of 384 considered barriers, it can be seen that barriers for Ni-vacancy interchanges are significantly higher than for Fe-vacancy interchanges, suggesting Ni atoms are significantly less mobile in these alloys.}
	\label{fig:total_barrier_histogram_fe_ni}
\end{figure}

Overall, it is determined that energetic barriers to lattice vacancy migration for Ni-vacancy interchanges are significantly higher than those for analogous Fe-vacancy interchanges in the Fe-Ni supercell configurations modelled here, in alignment with a previous \textit{ab initio} computational study~\cite{zhao_defect_2016} as well as with experimental outcomes~\cite{macewan_diffusion_1959, de_reca_self-diffusion_1967, million_diffusion_1981}. This finding is illustrated in Fig.~\ref{fig:total_barrier_histogram_fe_ni}, where histograms of the barriers associated with Ni-vacancy interchanges as compared to Fe-vacancy interchanges are presented.  Ni-vacancy interchanges clearly have consistently higher energy barriers than do their Fe-vacancy counterparts. This trend holds across the full range of considered compositions, as well as across the range of considered supercells with disordered, partially ordered, and fully ordered atomic arrangements. Across the dataset, the average barrier for a Ni-vacancy interchange $1.13 \pm 0.12$~eV, while for an Fe-vacancy interchange this number is 40\% smaller at $0.78 \pm 0.11$~eV. Given that, in accordance with the Arrhenius equation, diffusion coefficients (reaction rates) are exponentially related to barrier heights (activation energies), we { contend} that this substantial difference between average lattice vacancy migration barrier heights for { Ni-vacancy interchanges as compared to and Fe-vacancy interchanges} is { a key contributing factor to} the experimentally-observed differences in { atomic mobility} between the two species.

\begin{table}[t]
\caption{Calculated lattice vacancy migration barriers for the four possible atom-vacancy interchanges which occur in the pristine, atomically ordered, L1\textsubscript{0} phase of FeNi. Both interchanges involving Ni atoms have higher barriers than those involving Fe atoms. Additionally, interchanges between layers of atoms encounter different barriers to those within layers for both Fe and Ni atoms---indicative of the potential for anisotropic lattice vacancy diffusion coefficients in the L1\textsubscript{0} phase.}
\label{table:l10_phase_barriers}
\begin{ruledtabular}
\begin{tabular}{lcc}
             & \multicolumn{2}{c}{Vacancy location} \\
Atom interchanged & Fe layer          & Ni layer         \\ \hline
Fe           & 0.912~eV                   & 0.865~eV                  \\
Ni           & 1.315 eV                   & 1.024 eV                 
\end{tabular}
\end{ruledtabular}
\end{table}

In the case of the atomically ordered L1\textsubscript{0} phase containing a single, isolated vacancy, it is possible to identify two possible lattice positions for a vacancy--- either in the Fe layer or in the Ni layer. Continuing, vacancies in each of these two positions can participate in an additional two possible nearest-neighbour atomic interchanges---either within a layer or between layers. (All other atomic interchange possibilities are equivalent by symmetry.) The calculated lattice vacancy migration barriers for these four possibilities are shown in Table~\ref{table:l10_phase_barriers}. It can be seen that the two possible barriers where an Ni atom is interchanged with a vacancy have higher associated barrier heights than the two possible barriers where an Fe atom is interchanged with a vacancy, consistent with the findings for the ensemble of supercell configurations considered in Fig.~\ref{fig:total_barrier_histogram_fe_ni}. Furthermore, within the atomically ordered L1\textsubscript{0} phase, the calculated barrier heights for vacancy migration within a layer are different for those between layers. This effect is more pronounced for Ni than for Fe but is present for both chemical species. This result suggests that lattice vacancies in the L1\textsubscript{0} phase may exhibit a degree of anisotropy in their migration.

\subsection{Structural origins of differing barrier heights}
\label{sec:results_structural}

\begin{figure}[t]
    \centering
	\includegraphics[width=\linewidth]{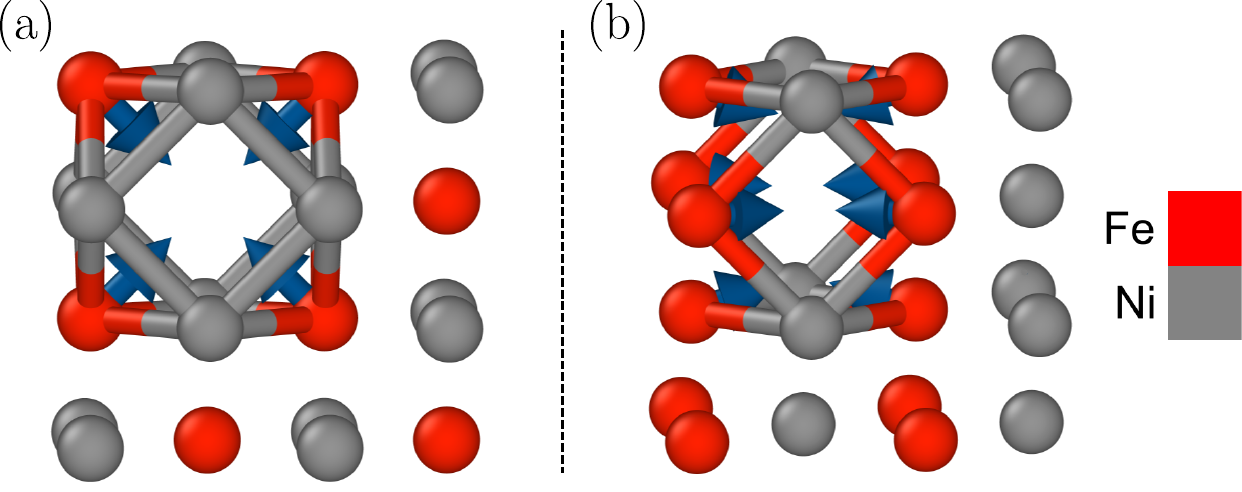}
	\caption{Diagram to show the effects of removing an atom from either (a) an Fe layer or (b) an Ni layer of the pristine, ordered L1\textsubscript{0} structure of equiatomic FeNi. The bonds highlighted are the Fe-Ni bonds that become shorter and the arrows show the direction in which atoms in the simulation cell move. Note that the structure shown in panel (b) is rotated compared to the structure shown in panel (a) for ease of viewing. Images generated using \textsc{Ovito}~\cite{stukowski_visualization_2010}.}\label{fig:vacancy_relaxation_schematic}
\end{figure}

The physical origin of the substantially higher lattice vacancy migration barriers calculated for Ni-vacancy interchanges compared to those for a Fe-vacancy interchanges may be understood via examination of how the alloy's atomic-scale structure is altered around lattice vacancies. We begin by inspecting the equiatomic, atomically ordered, L1\textsubscript{0} phase, where the only two distinct possibilities are for a lattice vacancy in either a Fe or a Ni layer. In this case, removal of an atom to create a vacancy in the cell causes local atomic relaxations around the vacancy, schematically shown in Fig.~\ref{fig:vacancy_relaxation_schematic}. As expected, the 12 atoms immediately neighbouring the vacancy experience the most significant deviations from their original atomic positions. However, we also find that Fe atoms relax into the vacancy significantly more than do Ni atoms. Across the two considered possible vacancy locations in the L1\textsubscript{0} phase, Fe atoms around the vacancy relax, on average, by 0.0428 \AA, while Ni atoms relax, on average, by 0.0127 \AA. (These deviations correspond, respectively, to 1.7\% and 0.5\% of the Fe-Ni nearest neighbour distance for the pristine L1\textsubscript{0} phase.) The directions in which these moves occur are shown in Fig.~\ref{fig:vacancy_relaxation_schematic}, demonstrating the pronounced Fe atom relaxation into the vacancy. 

\begin{figure}[t]
    \centering
	\includegraphics[width=\linewidth]{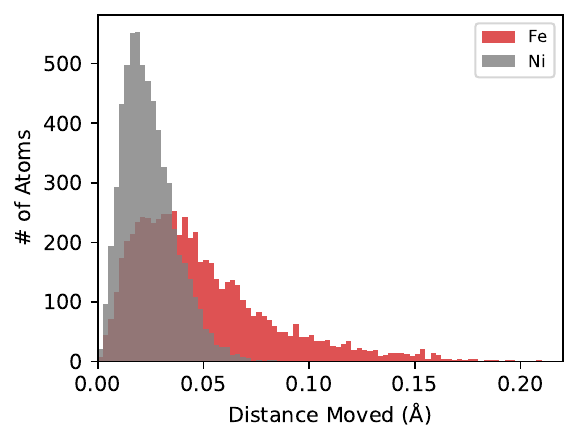}
	\caption{Histograms of the distance by which atoms move in vacancy-containing supercells during `relaxation' (geometry optimisation) following introduction of a vacancy. We compare results separately for Fe atoms (red) and Ni atoms (grey). It can be seen that Fe atoms move significantly more than Ni atoms. This is because Fe atoms `relax' into vacancies in our calculations, while Ni atoms remain more rigidly fixed to their lattice sites.}
	\label{fig:av_relaxation}
\end{figure}

More broadly, across our entire dataset of relaxed vacancy-containing supercells, a similar result holds.  Fig.~\ref{fig:av_relaxation} shows histograms of the average displacement of atoms in the simulation cell during a geometry optimization following removal of a single atom (selected at random) to create a vacancy. Fe atoms relax significantly more than do Ni atoms, with the average displacement of Fe atoms following creation of a vacancy of 0.015$\pm$0.004~\AA, whereas the average displacement of Ni atoms is less than half that value, at 0.007$\pm$0.002~\AA. We assert that this feature is the primary origin of the increased lattice vacancy migration barrier heights for Ni atoms in comparison with those for Fe atoms in this \textit{ab initio} study.

\begin{figure}[b]
    \centering
	\includegraphics[width=\linewidth]{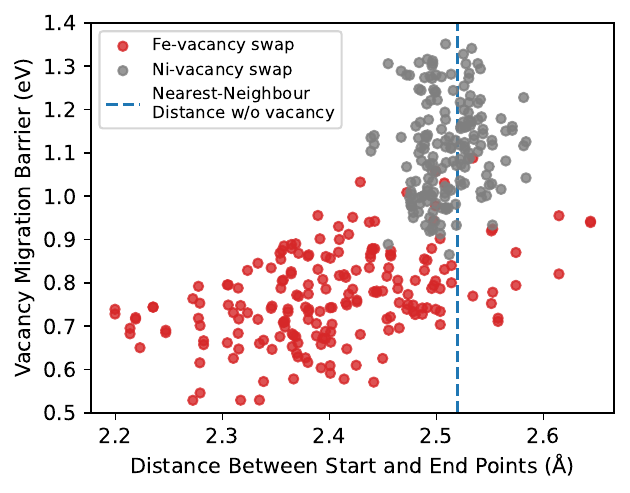}
	\caption{Lattice vacancy migration barrier height versus total distance between start and end points (initial and final NEB images) across the ensemble of ferromagnetic Fe-Ni supercells considered in this study, separated into barriers where an Fe atom moves and barriers where an Ni atom moves. The blue, vertical, dashed line indicates the average distance between nearest neighbour atoms in the considered simulation cells before a vacancy was introduced. It can be seen that the height of a vacancy migration barrier is positively correlated with the distance a given atom travels during the interchange.}
	\label{fig:total_dist}
\end{figure}

To further test this assertion, in Fig.~\ref{fig:total_dist} we plot barrier height against the total distance moved by the atom moving into the vacancy between initial and final NEB images. (These images correspond respectively to the location of this atom before the interchange begins, and the location of the atom after the interchange has occurred.) The Ni atoms, which relax by a comparatively small amount around vacancies, are clustered with large barrier heights so that the average distance travelled by Ni atoms is approximately equal to the nearest-neighbour distance in a simulation cell before the introduction of the vacancy.
The Fe atoms, on the other hand, show clear correlation between total distance moved and the barrier height: the smaller the distance moved between initial and final NEB images, the lower the barrier. These shorter total distances between initial and final images for Fe-vacancy interchanges originate in the aforementioned structural changes, where Fe atoms relax towards the vacancy in both the initial and final images. No such correlation is observed for the Ni atoms in our calculations.

\begin{figure}[b]
    \centering
	\includegraphics[width=\linewidth]{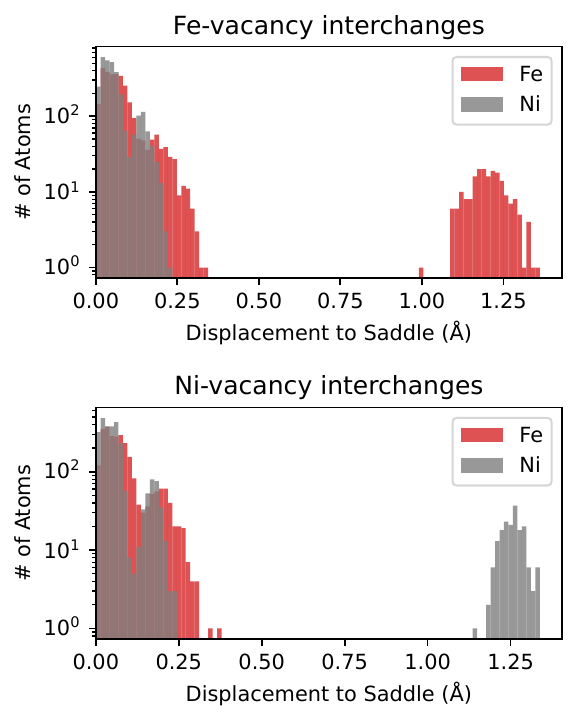}
	\caption{{ Distance moved by atoms in the simulation cell between the initial state of a NEB calculation and the saddle point for Fe-vacancy interchanges (top) and Ni-vacancy interchanges (bottom). The peaks corresponding to large displacements correspond to the atoms migrating. The bimodal nature of the distribution to the left of each plot relates to differing displacements for atoms in the immediate vicinity of the migration compared to those further away in the simulation cell. Note the use of a logarithmic scale on the vertical axis for both plots.}}\label{fig:saddle_point_environment}
\end{figure}

{
Finally, we consider the local environment of an atom at the saddle point during an atom/vacancy interchange. Shown in Fig.~\ref{fig:saddle_point_environment} are histograms of the distances moved by all atoms in the considered simulation cells between the initial (relaxed) vacancy-containing configuration and the saddle point. (Note that these data are plotted on a logarithmic scale.) The top panel shows results for Fe-vacancy interchanges, while the bottom panel shows results for Ni-vacancy interchanges. The peaks to the right of both plots---corresponding to large displacements---are those data points corresponding to the atoms migrating, while those peaks to the left of both plots correspond to other atoms in the simulation cell. 

Here, three key observations should be made. First is that---when considering the data relating to other atoms in the simulation cell---overall it can be seen that Fe atoms deviate more significantly from their starting positions than do Ni atoms, which is indicated by the broader width of the red Fe distribution as compared to the grey Ni distribution. This is broadly in accordance with the data shown in Fig.~\ref{fig:av_relaxation}, and is further evidence that Fe atoms are more able to move to accommodate changes to their local crystal environment than are Ni atoms. The second observation to be made relates to the bimodality of the distributions at low displacements. These have a clear interpretation in terms of the distance of atoms in the simulation cell from the migrating atom; those atoms directly neighbouring the migrating atom are displaced much more than those further away. The third observation relates to the splitting of the two peaks in the low-displacement distributions. Particularly for Ni atoms, those atoms immediately neighbouring the migrating atoms appear to be shifted by a greater amount when the migrating atom is a Ni atom as opposed to when it is an Fe atom. Overall, we conclude that Ni atoms remain more rigidly fixed to their starting lattice positions than do Fe atoms. It remains to elucidate the origin of this phenomenon.
}

\subsection{Correlation between spin polarised electronic structure and local lattice distortions around vacancies}
\label{sec:results_magnetic}

The dominant features of the spin-polarised electronic structure of a ferromagnetic, Fe-Ni alloy are its majority spin $d$-bands---which are completely filled, in line with Hund's Rules---and its partly filled minority spin $d$-bands.  Consequently, the majority-spin itinerant $d$-electrons behave very similarly around both Fe and Ni sites, whereas the disparate exchange and correlation effects associated with Fe and Ni atoms cause the minority-spin $d$-electrons to behave in a different manner.  A large spin polarization is therefore sustained around each Fe site, which manifests as a local moment~\cite{gyorffy_first-principles_1985}, in contrast with a much smaller more itinerant spin polarisation in the vicinity of each Ni site~\cite{staunton_using_2014}. This minority-spin electron behaviour is also understood to be the main driver for the transition of the A1 phase to the L1\textsubscript{0} phase at low enough temperatures~\cite{woodgate_integrated_2024}.

When a vacancy is introduced into the Fe-Ni lattice, the local electron density is altered and the atoms shift their positions to minimize the total energy. We contend that the differing number of $d$-electrons associated with Fe and Ni atoms, the associated differing exchange-correlation splitting of the majority and minority spin electron states around atoms, together with the constraint of filled majority spin $d$-states, produce a combined effect which is instrumental in shaping the atomic-scale energy landscape of the alloy.

In terms of a tight-binding picture, the electronic structure of a transition metal alloy is set up by $d$-orbitals centred on atomic sites, the lobes of which point towards various surrounding atoms, and the relative energy of different orbitals (determining the order in which they fill) is determined primarily by electrostatic effects and the local crystallographic environment. In the fcc A1 phase of the alloys considered in this work, for example, all majority spin $d$-states associated with both Fe and Ni are occupied, so it is consideration of the filling of the minority spin $d$-states which is a crucial part of the analysis of the electronic effects dictating the stability of the alloy phase. In the minority-spin channel for this fcc phase, the triplet $t_{2g}$ states lie at lower energies than the doublet $e_{g}$ states. For Fe atoms, the $e_{g}$ states are largely unfilled, and the $t_{2g}$ states---which are partially filled---sit at energies close to the Fermi energy of the material. By contrast, for Ni atoms, the minority-spin $t_{2g}$ states are largely filled, and it is the partially-filled $e_{g}$ states which sit at energies close to the Fermi energy.

In real-space, these $t_{2g}$ states are those $d$-orbitals with lobes pointing directly at nearest-neighbour atoms which form broad $d$-bands and contribute strongly to bonding. By contrast, the $e_{g}$ states are those $d$ orbitals with lobes which point towards \textit{second}-nearest neighbour atoms of a given atom forming narrower bands and smaller bonding contribution. When a vacancy is introduced, the local symmetry of the system is lowered and the ions and electrons rearrange themselves to find a new arrangement which minimises the system's total energy. We suggest that Ni atoms, which have their $t_{2g}$ states placed well below the Fermi energy, have significantly less scope to rearrange themselves in the vacancy-containing structure than Fe atoms. The partially-filled $d$ states associated with these Fe atoms can readily form new linear combinations of orbitals which  enable the Fe atoms to move further into the newly-created space, minimising the total energy of the system. In turn, in this more open local environment, the Fe atoms exhibit increased exchange splitting which supports a marginally increased magnetic moment as compared to the pristine bulk.

In summary, when a vacancy is introduced, the energy is lowered by a Fe atom close to it repositioning to optimise the bonding it has with its nearest Fe and Ni neighbours. Such an effect is much smaller for the Ni atoms given their electronic structure around the Fermi level. It is energetically favourable for Fe atoms to move in on the open space created by a vacancy while increasing their associated magnetic moments. In contrast it is energetically preferable for Ni atoms to remain `closer' to the original bulk positions. The results are the relatively high barriers for Ni-vacancy interchanges compared to those of the Fe-vacancy ones.

\begin{table*}[t]
\caption{Values of the Fe and Ni magnetic moments before and after a vacancy is introduced into either a Fe or Ni layer of the atomically ordered L1\textsubscript{0} phase of FeNi. Magnetic moments on Fe atoms grow more than those on Ni atoms when a neighbouring vacancy is introduced.}
\label{table:increase_spin}
\begin{ruledtabular}
\begin{tabular}{lcccc}
             & \multicolumn{2}{c}{Vacancy in Fe layer} & \multicolumn{2}{c}{Vacancy in Ni layer}\\
 & Fe atoms nearest vacancy          & Ni atoms nearest vacancy & Fe atoms nearest vacancy          & Ni atoms nearest vacancy        \\ \hline
Before           &  2.805~$\mu_B$ & 0.451~$\mu_B$& 2.805~$\mu_B$&    0.451~$\mu_B$           \\
After          & 2.937~$\mu_B$  & 0.472~$\mu_B$& 2.845~$\mu_B$  & 0.465~$\mu_B$                
\end{tabular}
\end{ruledtabular}
\end{table*}

\begin{figure}[t]
    \centering
	\includegraphics[width=\linewidth]{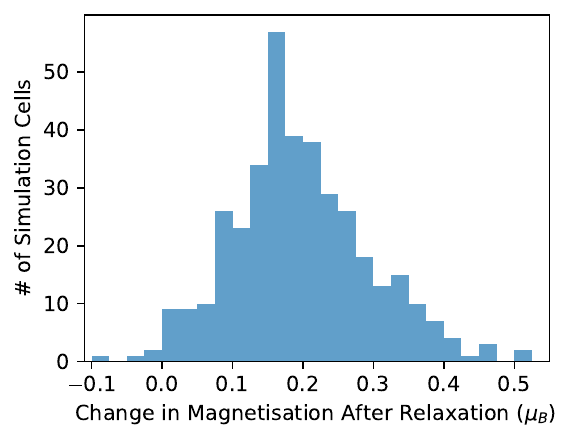}
	\caption{Histogram of the change in the total magnetisation of a simulation cell between the start and end of the geometry optimization procedure following creation of a lattice vacancy. Across the ensemble of considered supercells, there is an average increase of in total supercell magnetization of $0.2 \pm 0.1$~$\mu_B$, which originates primarily from Fe atoms around lattice vacancies. This result suggests magnetism and spin-polarized electronic structure plays a key role in determining the energy landscape around such point defects.}\label{fig:Spin_increase_all_barriers}
\end{figure}

This assertion is tested by re-examination of the pristine, atomically ordered, L1\textsubscript{0} phase, where a lattice vacancy occurs in one of two distinct locations: in either a Fe or Ni layer. Table~\ref{table:increase_spin} shows the calculated atomic magnetic moments for both Fe and Ni atoms before and after a vacancy is introduced and indicates that the change in magnetic moment associated with Fe atoms is larger than the change associated with Ni atoms, by approximately a factor of five for both possible vacancy locations.

Moreover, this proposed mechanism accounts for the trends observed across the ensemble of supercell calculations as evidenced by  Fig.~\ref{fig:Spin_increase_all_barriers}. The figure shows a histogram of the change in total magnetization of a given vacancy-containing supercell between the initial configuration (when a vacancy is newly created) and the final, optimized configuration (when local distortions around the vacancy have been realized). During the geometry optimization, the total magnetization of simulation cells grows (on average) by $0.2 \pm 0.1$ $\mu_B$, accruing mainly on the Fe atoms neighbouring vacancies. 

\subsection{Importance of the local atomic environment}

It is important to examine the impact of the local atomic environment when assessing barriers to lattice vacancy migration for different chemical species in an alloy such as the Fe-Ni alloys considered in this work. This aspect may be investigated by examining an ensemble of structures including fully- and partially atomically ordered structures, as well as structures representative of the disordered phase. This is particularly relevant in the Fe-Ni alloys considered in this work, as it is known from previous computational studies that the atomically ordered L1\textsubscript{0} phase is marginally elastically stiffer than the atomically disordered A1 phase~\cite{tian_density_2019, woodgate_integrated_2024}. (For example, Ref.~\cite{woodgate_integrated_2024} reports a DFT-calculated bulk modulus of the L1\textsubscript{0} phase of 189.7~GPa, while for the A1 phase the reported bulk modulus is 180.2~GPa.) We infer, therefore, that the crystal lattice of the ordered L1$_0$ phase is marginally more rigid than that of the disordered A1 phase, and that the rigidity of atomic bonds may differ slightly as well. It is therefore necessary to inspect how lattice vacancy migration barriers differ between disordered and ordered phases.

\begin{figure}[t]
    \centering
	\includegraphics[width=\linewidth]{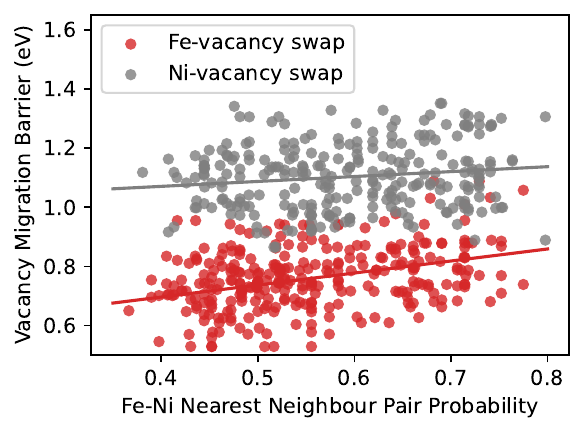}
    \caption{Scatter plot of the conditional probability of Fe-Ni atoms being found as nearest-neighbour pairs in a given simulation cell versus the barrier height obtained upon removing a selected atom from the simulation cell and performing a NEB calculation. The solid grey and red lines indicate linear lines of best fit for Ni atoms and Fe atoms, respectively, and are to guide the eye. For both chemical species, there is comparatively weak correlation between the local chemical environment (quantifying atomic short- and long-range order) and the calculated barrier height.}\label{fig:barrier_height_asro}
\end{figure}

To quantify the probability of finding Fe-Ni pairs as nearest neighbours in a given simulation cell, we consider the probability of finding an Fe-Ni atom pair at nearest-neighbour distance. Such data naturally incorporate both varying composition as well as varying degrees of atomic order \footnote{Note that we calculate these probabilities \textit{before} an atom is removed from the cell to create a lattice vacancy, and take an average across the whole simulation cell, with periodic boundary conditions applied using the minimum image convention as implemented in ASE~\cite{hjorth_larsen_atomic_2017}.}. For example, at the equiatomic composition, the atomically disordered A1 phase is characterized by a probability of 0.5 of Fe-Ni pairs occurring as nearest-neighbours, while the pristine L1\textsubscript{0} phase is characterized by a probability of finding Fe-Ni pairs as nearest-neighbours of exactly 2/3, as 8/12 nearest neighbours are anticipated to be unlike pairs in this structure. The magnitude of the energetic barriers to atom-vacancy interchange versus the probability of finding a Ni atom neighbouring an Fe atom in the simulation cell is depicted in Figure~\ref{fig:barrier_height_asro}. It can be seen that for both Ni-vacancy interchanges (represented by the grey data markers), as well as for Fe-vacancy interchanges (represented by the red data markers) there are positive correlations between the pair probability data and predicted lattice vacancy migration barrier height. In essence, simulation cells with an increased number of Fe-Ni nearest-neighbour pairs have marginally higher barrier heights than those with fewer Fe-Ni pairs. We attribute this to the relative rigidity of Ni atomic bonds in our simulations, which result in Ni atoms not deviating much from their ideal crystallographic positions in both A1 and L1\textsubscript{0} structures, limiting the mobility of Fe atoms in cells where an increased number of Fe-Ni pairs are found.

{
\subsection{Lattice vacancy formation energies}

\begin{table}[b]
{
\caption{Vacancy formation energies (enthalpies) for Fe and Ni vacancies in both disordered fcc (A1) FeNi and ordered tetragonal (L1\textsubscript{0}) FeNi. For Fe vacancies, values are provided with both the energy per atom of bcc Fe and fcc Fe as a reference state determining the chemical potential.}\label{table:vacancy_formation_energies}
\begin{ruledtabular}
\begin{tabular}{lccc}
Phase    & bcc Fe $E^\textrm{f}_\textrm{v}(\textrm{Fe})$ & fcc Fe $E^\textrm{f}_\textrm{v}(\textrm{Fe})$ & $E^\textrm{f}_\textrm{v}(\textrm{Ni})$ \\ \hline
A1  & $1.65 \pm 0.10$~eV & $1.81 \pm 0.10$~eV & $1.87 \pm 0.06$~eV\\
L1\textsubscript{0} & 1.86 eV & 2.02 eV & 2.10 eV
\end{tabular}
\end{ruledtabular}
}
\end{table}

Of course, lattice vacancy migration barriers alone do not solely determine the rates of species-dependent atomic {mobility} in alloys, and it is also important to consider the (free) energetic cost of vacancy formation. To address this aspect, we consider explicitly vacancy formation energies (enthalpies) for both Fe and Ni atoms in both the atomically disordered (A1) phase, as well as the atomically ordered L1$_0$ phase. Our calculated vacancy formation energies (enthalpies) are provided in Table~\ref{table:vacancy_formation_energies}. Our calculated values for the disordered A1 phase compare favourably to those calculated by Zhao \textit{et al.}~\cite{zhao_defect_2016} using the Widom insertion method~\cite{widom_topics_1963} to determine the chemical potential, though we do not distinguish between the various possible chemical environments as those authors do. Our numbers are also, on the whole, marginally higher than theirs, which we understand as originating from our high plane wave cutoff and reduced electronic smearing compared to their calculations.

When the vacancy formation energies of Table~\ref{table:vacancy_formation_energies} are considered, and compared with the vacancy migration barriers presented earlier in the work, e.g. Table~\ref{table:l10_phase_barriers},  Figure~\ref{fig:total_barrier_histogram_fe_ni}, we see that the energetic cost of vacancy formation is significantly higher than that of vacancy migration. Additionally, in a close-packed metallic system---such as the fcc Fe-Ni alloys considered in this work---creation of self-interstitials (Frenkel pairs) is energetically highly unfavourable and vacancies are understood instead to originate from the surface and propagate inwards. We therefore { contend} that it is the difference in average lattice vacancy migration barrier for Fe-vacancy interchanges as opposed to Ni-vacancy interchanges which dominates the experimentally observed differences in atomic mobility between the two species.
}

\section{Summary and Conclusions}
\label{sec:conclusions}

In summary, results have been presented examining lattice vacancy migration barrier heights in near-equiatomic, ferromagnetic, Fe-Ni alloys within a first-principles electronic structure framework, where barrier heights were accurately determined using the nudged elastic band method. Across an ensemble of near-equiatomic, fcc Fe-Ni alloy supercells spanning a range of Fe:Ni ratios and degrees of atomic short- and long-range order, the average lattice vacancy migration barrier height for for a Ni atom interchanging with a lattice vacancy ($1.13\pm 0.12$~eV) was found to be 40\% higher than that for an Fe atom interchanging with a lattice vacancy ($0.78\pm 0.11$~eV). 

This key result motivated examination of local structural distortions in the atomic environment surrounding lattice vacancies to conclude that Fe atoms structurally relax into vacancies significantly more than do Ni atoms; in contrast, the Ni atoms remain rigidly located at their original lattice position that was determined before a vacancy was introduced. This difference in structural relaxation behaviour is understood to be of magnetic origin, with the Fe atoms that are able to `relax into' lattice vacancies experiencing a growth in magnetic moment owing to the more `open' local environment, and with a consequent reduction in energy. That Fe atoms moving into the space created by lattice vacancies causes the significant difference in lattice vacancy migration barrier heights is supported by the noted correlation between distance travelled by a migrating atom during the atom-vacancy interchange (defined in Sec.~\ref{sec:methods}) and the energetic magnitude of the barrier height. In our dataset, those atoms which move a long way during an atom-vacancy interchange have higher barriers than those which move only a short distance between initial and final NEB images.

These results provide atomic-scale insight into the long-standing, experimentally established observation of highly sluggish diffusion of Ni atoms in Fe-Ni alloys as well as { evidencing a likely major contributing mechanism driving this behaviour}, with implications both for the formation of hard magnetic L1\textsubscript{0} FeNi (tetrataenite) as well as for Fe-Ni-based alloys more generally. Future work will examine how external factors such as the application of external forces, including pressure, strain, and/or applied magnetic field, influence these energetic lattice vacancy migration barriers. Moreover, our established dataset~\cite{dataset} of almost 200 highly accurate NEB calculations, performed across a range of Fe-Ni alloy compositions and containing structures with varying degrees of atomic order, provides a very high-quality training and/or validation set for use in the development of a machine-learned interatomic potential (MLIP) suitable for the study of atomic diffusion in this class of alloys. Such an MLIP is imperative in this context, as previous works have shown~\cite{zhao_defect_2016, fisher_first_2024} that empirical and semi-empirical potentials fail to capture lattice vacancy migration barrier heights with acceptable accuracy. We are currently in the process of training such a potential for Fe-Ni alloys for general metallurgical applications. The first use case of this potential will be in kinetic Monte Carlo (kMC) simulations, which will determine temperature-dependent atomic diffusion coefficients and produce further insight into the sluggish kinetics of the well-studied A1/L1\textsubscript{0} disorder-order transition in Fe-Ni alloys.

\section*{Acknowledgments}
The authors would like to thank Dr Albert Bart\'ok-P\'artay (University of Warwick, UK) { and Dr Philip Hasnip (University of York, UK)} for helpful discussions. This work was supported by the UK Engineering and Physical Sciences Research Council (EPSRC), Grant EP/W021331/1, by the U.S. Department of Energy, Office of Basic Energy Sciences under Award Number DE SC0022168 (for atomic insight) and by the U.S. National Science Foundation under Award ID 2118164 (for advanced manufacturing aspects). C.D.W. acknowledges support from an EPSRC Doctoral Prize Fellowship at the University of Bristol, Grant EP/W524414/1. 

The authors also acknowledge the use of compute resources provided by the Isambard 3 high-performance computing (HPC) facility. Isambard 3 is hosted by the University of Bristol and operated by the GW4 Alliance (\href{https://gw4.ac.uk}{https://gw4.ac.uk}) and is funded by UK Research and Innovation; and the EPSRC, Grant EP/X039137/1. Additional compute resources were provided by the Scientific Computing Research Technology Platform (SCRTP) of the University of Warwick and by the Advanced Computing Research Centre (ACRC) of the University of Bristol.

A.F.\ and J.B.S.\ designed the research, with input from C.D.W., X.Z., G.C.H., and L.H.L.. C.D.W.\ performed the KKR-CPA calculations and Monte Carlo simulations to generate the alloy supercells. A.F.\ carried out the majority of the DFT and NEB calculations, with support from C.D.W.. { Calculations of vacancy formation energies were performed by C.D.W..} The project was supervised by J.B.S.\ and L.H.L.. A.F.\ and C.D.W.\ jointly prepared the first draft of the manuscript. Subsequently, all authors revised the manuscript and approved its final version.

{

\section*{Data Availability Statement}
The data that support the findings of this article are openly available~\cite{dataset}.
}
\appendix

{
\section{Supercell generation}
\label{appendix:supercell}

As discussed briefly in Sec.~\ref{subsec:supercell_generation}, in this work we follow the approach outlined by Shenoy \textit{et al.}~\cite{shenoy_collinear-spin_2024} to produce alloy simulation cells with varying degrees of atomic order. This means performing fixed lattice Metropolis Monte Carlo simulations using atom-atom effective pair interactions calculated within an effective medium theory. Equilibrated configurations drawn from simulations performed at low temperatures will have a high degree of atomic ordering, while those configurations equilibrated at high temperatures will be representative of the atomically disordered phase. In this Appendix, we outline this approach in detail.

An alloy configuration is described by a set of lattice site occupancies, $\{ \xi_{i\alpha}\}$, where $\xi_{i\alpha}=1$ if site $i$ is occupied by an atom of chemical species $\alpha$, and 0 otherwise. The internal energy associated with a particular configuration is then described by a simple pairwise Bragg-Williams Hamiltonian~\cite{bragg_effect_1934},
\begin{equation}
    H = \sum_{i \alpha; j \alpha'} V_{i \alpha; j \alpha'} \xi_{i\alpha} \xi_{j \alpha'},
    \label{eq:bragg-williams}
\end{equation}
where the atom-atom \textit{effective pair interaction} (EPI), $V_{i \alpha; j \alpha'}$, describes the energy associated with an atom of chemical species $\alpha$ on lattice site $i$ interacting with an atom of chemical species $\alpha'$ on lattice site $j$. We assume these interchange parameters are isotropic, homogeneous, and have finite range, which simplifies Eq.~\ref{eq:bragg-williams}. In this work, the EPIs, $\{V_{i \alpha; j \alpha'}\}$, are obtained using the alloy $S^{(2)}$ theory~\cite{khan_statistical_2016, woodgate_compositional_2022, woodgate_modelling_2024}. This approach uses the language of concentration waves~\cite{gyorffy_concentration_1983} to describe atomic-scale chemical fluctuations in a solid solution. The theory uses the Korringa--Kohn--Rostoker (KKR)~\cite{korringa_calculation_1947, kohn_solution_1954, ebert_calculating_2011} formulation of DFT, with substitutional chemical disorder described via the coherent potential approximation (CPA)~\cite{soven_coherent-potential_1967, gyorffy_coherent-potential_1972, stocks_complete_1978}, producing an effective medium representing the electronic structure of the disordered alloy. 

In this work, the \textsc{Hutsepot} package~\cite{hoffmann_magnetic_2020} was utilised to construct the self-consistent, one-electron potentials of DFT within the KKR-CPA formalism. Throughout, the alloy was simulated in a ferromagnetic state, as it is understood that magnetic order contributes substantially to the thermodynamic stability of the atomically ordered L1\textsubscript{0} phase~\cite{izardar_interplay_2020, ruban_qualitative_2024, woodgate_integrated_2024}. Proceeding, the concentration wave analysis was performed and atom-atom EPIs fitted to the first four coordination shells of the fcc lattice ~\cite{khan_statistical_2016, woodgate_compositional_2022, woodgate_modelling_2024}. These fitted EPIs are available in the open-access dataset associated with this work~\cite{dataset}. As outlined in the main text, the Metropolis Monte Carlo simulations, performed with the \textsc{Brawl} package~\cite{naguszewski_brawl_2025}, used 32-atom simulation cells (supercells) with periodic boundary conditions applied in all three directions. A given simulation cell is initially populated at random with overall atomic concentrations consistent with the desired chemical composition was then equilibrated at the desired temperature using a run of 32,000 trial Metropolis-Hastings atomic interchanges, \textit{i.e.} 1000 trial Monte Carlo moves per lattice site. 

To cover both stoichiometric FeNi and off-stoichiometric compositions, we used five distinct compositions when constructing our Fe$_x$Ni$_{1-x}$ supercells. The $x$ values used (to 2 decimal places) were $x= 0.41$, 0.44, 0.50, 0.56, and 0.59, corresponding to 13, 14, 16, 18, and 19 Fe atoms in the simulation cell. For each composition, eight equilibrated samples were drawn at simulation temperatures of 300, 400, 600, 1200, and 3000 K. (The A1--L1\textsubscript{0} transition temperature of the pairwise model utilised in this work has previously been found to be approximately 500 K for the equiatomic composition~\cite{woodgate_integrated_2024}, so the equilibration temperatures were chosen to sample the alloy in atomically ordered, partially ordered, and disordered states.) Specifically, the samples drawn from high-temperature simulations will have little to no atomic short-range order, those drawn at intermediate temperatures have a degree of short-range order present, and those drawn at low temperatures will be representative of the alloy in a near perfectly ordered state.}

\end{document}